\documentclass[twocolumn,openany,openright,amsmath,amssymb,superscriptaddress,prb]{revtex4-2}

\usepackage{float}
\usepackage{latexsym} 
\usepackage{amsmath,amsthm}
\usepackage{ifpdf}
\usepackage{epstopdf}
\usepackage{soul}
\usepackage{dcolumn}
\usepackage{bm}
\usepackage{braket}
\usepackage{wrapfig}
\usepackage{comment}
\usepackage[abs]{overpic}
\usepackage{graphicx,graphics}
\usepackage{dcolumn}
\usepackage{wasysym}
\usepackage{latexsym,verbatim}
\usepackage{color}
\usepackage[framemethod=default]{mdframed}
\usepackage[breaklinks=true,colorlinks,citecolor=blue,linkcolor=blue,urlcolor=blue]{hyperref}
\usepackage[makeroom]{cancel}
\usepackage{fancybox}
 \usepackage{tikz-feynman}

\DeclareMathOperator{\rank}{rank}

\newcommand{\diag}{\textrm{diag}}

\newcommand{\qq}{\mathbf{q}}

\begin{document}
\def \k{\bm k}
\def \l{\bm l}
\def \i{\bm i}
\def \p{\bm p}
\def \ppi{\bm \pi}
\def \q{\bm q}
\def \r{\bm r}
\def \s{\bm s}
\def \P{\bm P}
\def \R{\bm R}
\def \qq{\bm q}
\def \A{\bm A}
\def \D{\bm D}
\def \beq{\begin{equation}}
\def \eeq{\end{equation}}
\def \beal{\begin{aligned}}
\def \eal{\end{aligned}}
\def \bes{\begin{split}}
\def \ees{\end{split}}
\def \besu{\begin{subequations}}
\def \esu{\end{subequations}}
\def \g{\gamma}
\def \G{\Gamma}
\def \ac{\alpha_c}
\def \barr{\begin{eqnarray}}
\def \earr{\end{eqnarray}}

\title{The Fracton quantum gas}
\author{Leone Di Mauro Villari}
\author{Alessandro Principi}
\affiliation{Department of Physics and Astronomy, University of Manchester, Manchester M13 9PL, UK}
\begin{abstract}
Starting from a simple dipole-conserving Hamiltonian model, after introducing a proper quantisation framework, we construct the linear-response theory of a fracton quantum gas. We show how to consistently construct the density operator and we calculate the density-density linear-response function in the limit of large number of (fracton) flavours. We show how to construct its random-phase approximation, from which we calculate the \emph{fracton}-plasmon dispersion.
\end{abstract}
\maketitle
\section{Introduction}
One of the most challenging, but at the same time inspiring, pursue of condensed matter physics is the search for new states of matter. This endeavour has focussed on both traditional methods, such as the design of phase transitions and band structures \cite{nhu,pcx}, and more modern frameworks encompassing topology and non-equilibrium physics. In recent years there has been a growing interest in states of matter whose quantum excitations feature highly constrained motion. These excitation are called fractons. They either cannot move without creating additional excitations,
or they can only move in certain directions. Hitherto, two main approaches have been used in the description of the fracton phase of matter. The first, introduced in Chamon’s foundational paper \cite{cc} and other early works \cite{blt,cca,Haah,bhj,vhf,mwr} uses tools from quantum information theory to study exactly solvable spin and Majorana models. All these works focus on the interesting problem of quantum glassiness as they involve quantum many-body Hamiltonians, with local interactions and no quenched disorder, that are unable to reach their ground states as the environment temperature is lowered to absolute zero. These systems also have interesting applications to the problem of self-correction of quantum memories. In particular the immobility of fractons may lead to the development of robust quantum memories \cite{Haah,bhj,mwr}. Fractons have since been shown to arise in various other contexts ranging from topological crystalline defects \cite{pr2} to  plaquette-ordered paramagnets \cite{pms}.  The second approach and currently widely used \cite{pnr,pr,pr1,pr2,glo}, is centred on the study of certain tensor gauge theories. In this framework it is possible to derive the so called fracton hydrodynamics \cite{gln} and the so called fracton-elasticity duality \cite{pr2}, both of which provide an effective field theory for this novel state of matter. Since the structure of hydrodynamics is sensitive only on local symmetries it provides a general picture of fracton physics as multiple fracton phases can fall into the same hydrodynamic universality class. Furthermore the approach based on gauge theories is also the natural framework to study how fractons couple with external gauge fields. In this way, for example it is possible to show that fracton hydrodynamics
is described by a set of unusual conservation laws involving higher rank current operators. In certain fracton models the restricted mobility can, in fact, be explained as a consequence of the dipole (or higher multipole) moment conservation \cite{gromov,HY}.

In this paper we consider a very simple model which conserves the dipole moment, and as such allows for fracton excitations. It is  translational invariant and has been known since 1950 in nuclear physics \cite{kre,pal}. It is called the translational invariant shell model (TISM). A modified version of this model, with only nearest neighbors interactions, has been studied recently as a model for nonlinear hydrodynamics in fracton fluids in one dimension \cite{glo}. This model has been studied in the context of the quantum Hall regime \cite{simon}. The TISM is also the starting point of a recent study of the classical mechanics of non-relativistic fractons \cite{pgs}. After briefly reviewing the properties of the model Hamiltonian we define the density operator and construct the fracton linear response theory. We calculate it in the limit of large number of (fracton) flavours, thus constructing a random-phase approximation. We find an unusual scaling of the RPA dielectric function at low momentum which can be related to the restricted mobility, and calculate the the collective modes, which we term ``fracton-plasmons''. 
\section{The model}
The simplest possible Hamiltonian of a translationally-invariant dipole-moment-conserving many-particle system is (we set $\hbar=c=m=1$)
\begin{equation} \label{eq1}
{\hat H} = \frac{1}{2N} \sum_{i \neq j}^N [({\hat \p}_i - {\hat \p}_j)^2 + V( {\hat \r}_i- {\hat \r}_j)].
\end{equation}
where $\hat \p_i=(\hat p_{i,1},\dots,\hat p_{i,D})$, $\hat \r_i=(\hat r_{i,1},\dots,\hat r_{i,D})$  and $D=1,2,3$ is the dimensionality of the system. The Hamiltonian in Eq.~(\ref{eq1}) preserves the value of the center of mass (COM) position (or total dipole of the system), ${\hat \D} = N^{-1} \sum_i {\hat \r}_i$, and of the COM momentum, ${\hat \P}_{\rm cm} = \sum_i {\hat \p}_i$, {\it i.e.} $[{\hat H}, {\hat \D}] = [{\hat H}, {\hat \P}_{\rm cm}] = 0$. Here, ${\hat \r}_i$ and ${\hat \p}_i$ are the position and canonical momentum operators of the quantum particles (we denote their eigenvalues without the hat). The dipole and COM momentum obey the multipole algebra~\cite{glo} $[{\hat \D},{\hat \P}_{\rm cm}]=Q$ where $Q$ is the total charge (equal to the total number of particles, $N$).
%
The Hamiltonian~(\ref{eq1}) is equivalent to the ``translation invariant shell model'' \cite{kre} if $V({\hat \r}_i-{\hat \r}_j)$ is a harmonic potential,
which is of importance in nuclear physics~\cite{pal}. The equivalence with this model is useful as it allows to quantise this system as a collection of $N-1$ quantum particles,
as reviewed in the appendices \ref{AppA} and \ref{AppB}.

In adding external fields, we relax the conservation of the COM momentum, but continue to enforce the conservation of its position (and therefore of the total dipole). For example, we couple the system to a uniform external electric field ${\bm E}$ via a term of the form ${\bm E}\cdot{\hat {\bm D}}$. Such term ensures that $\partial_t {\hat {\bm D}} = i [{\hat H}, {\hat {\bm D}}] = 0$, while the COM momentum evolves according to $\partial_t {\hat \P}_{\rm cm} = i [{\hat H}, {\hat \P}_{\rm cm}] = {\bm E}$. At this point, a reader may be confused about how conserving the COM position, but not the COM momentum may even be possible. The solution of this puzzle is that canonical (${\hat {\bm p}}_i$) and physical (${\hat {\bm \pi}}_i$) momenta do not coincide. In fact, the kinetic part of the Hamiltonian in Eq.~(\ref{eq1}) can be rewritten as
\begin{equation} \label{eq4}
    {\hat H}_{\rm kin} = \frac{1}{2} \sum_{i=1}^N ({\hat \p}_i - N^{-1} {\hat \P}_{\rm cm})^2,
\end{equation}
which makes it clear that the physical momentum is given by ${\hat {\bm \pi}}_i = [{\hat {\bm r}}_i, {\hat H}] = {\hat \p}_i - N^{-1} {\hat \P}_{\rm cm}$. Using this definition, it is immediate to see that (i) ${\hat {\bm \Pi}}_{\rm cm} \equiv \sum_{i=1}^N {\hat {\bm \pi}}_i = 0$ and (ii) $\partial_t {\hat {\bm \Pi}}_{\rm cm} = 0$ (note indeed that $\partial_t {\hat \p}_i = {\bm E}$).

Equation~(\ref{eq4}) suggests also that, for a system of free particles [$V( {\hat \r}_i- {\hat \r}_j) = 0$], eigenstates should be labeled in terms of the eigenvalues of the operators ${\hat {\bm \pi}}_i$.
A generic state $|\psi_\pi \rangle = |{\bm \pi}_1, \ldots, {\bm \pi}_N\rangle$ is such that ${\hat {\bm \pi}}_i|\psi_\pi\rangle = {\bm \pi}_i |\psi_\pi\rangle$, and its energy is $\sum_{i=1}^N {\bm \pi}_i^2/2$. 
As it is usually the case in many-body physics, the action of operators is expressed on the basis of non-interacting states, in this case $|\psi_\pi \rangle$. We start by defining the action of the density operator whose Fourier transform reads $n_{\bm q} = \sum_{i=1}^N e^{-i \q\cdot {\hat \r}_i}$. Using that the commutator with the physical momenta is $[{\hat {\bm \pi}}_i, e^{-i{\bm q}\cdot {\hat {\bm r}}_j}] = {\bm q} e^{-i{\bm q}\cdot {\hat {\bm r}}_j} (\delta_{ij} - N^{-1})$, one can show that $e^{-i \q\cdot {\hat \r}_i} |\psi_\pi\rangle$ is a state in which ${\bm \pi}_i \to {\bm \pi}_i + (N-1){\bm q}/N$ while all other ${\bm \pi}_j \to {\bm \pi}_j  - {\bm q}/N$ (for $j\neq i$). Thus, the action of the density on the states $|\psi_\pi \rangle$ preserves the COM (physical) momentum and dipole. Though natural, it is quite undesirable to work with these states to develop a linear response theory of the fracton gas, since the action of any operator involves all $N$ particles in the system, in other words the dynamics is \emph{non-local}. A way to circumvent this issue is to consider the thermodynamic limit ($N \to \infty$). We note that this is also the natural limit to study the linear response\cite{gv}. In this way the commutation relation for the physical momenta relation reduces to the standard one $[\ppi_i,\r_j]=\delta_{ij}$ and we can safely assume $\ppi_i\approx \p_i$. This solves the locality problem, but introduces an over-simplification. The TISM in the infinite volume limit reduces to a standard quantum gas ($\hat H \approx \sum_i \p_i^2/2$). It seems we have reached an impasse, either we deal with a non-local response or we lose the fracton behaviour of the system. To overcome this new problem we have to introduce another constraint that imposes locality and at the same time re-enforces the \emph{fractonic} nature of the TISM in the $N \to \infty$ limit, \emph{i.e} the conservation of the COM. This means that the action of the density operator on the states $|\psi_p \rangle = |{\bm p}_1, \ldots, {\bm p}_N\rangle$ 
becomes
%
%
\beq \label{eq:n_constraint}
 {\hat n}_{\bm q} |\psi_p \rangle = \frac{1}{N} \sum_{i\neq j} \ket{\p_1,\dots,\p_i+\q,\dots, \p_j-\q,\ldots,\p_N}.
\eeq
In this way the action of the density operator is equivalent to that of the {\it two-particle} density $n_{2,\q} = \sum_{i \neq j} e^{i\q\cdot ({\hat \r}_i- {\hat \r}_j)}$.
%
%
In second quantization~\cite{gv} 
%
\begin{equation}
\hat n_{2,\q} =  \sum_{\p_1\p_2}c^\dagger_{\p_1+\q}c^\dagger_{\p_2-\q}c_{\p_1}c_{\p_2} = :  {\hat n}_{\bm q}  {\hat n}_{-{\bm q}} :,
\end{equation}
where $:\ldots:$ stands for the normal ordering of operators with respect to the vacuum.

\section{Linear Response theory}
We now develop the linear-response theory for this system. The causal fracton density-density response function is given by~\cite{gv} $\theta(t)\braket{[n_{\q}(t),n_{-\q}(0)]}$, where $\theta(t)$ is the Heaviside step function. It is easy to show that, expanding this expression in terms of the eigenstates $|\psi_p \rangle$, and implementing the action of the density operator in Eq.~(\ref{eq:n_constraint}), the density-density response function is identical to
\beq \label{eq33}
\chi_{n_2 n_2}(\q,t) = \theta(t)\braket{[n^{(2)}_{\q}(t),n^{(2)}_{-\q}(0)]}. 
\eeq
This striking result, that the density and the pair-density response functions coincide is a consequence of the constraint imposed by the fracton nature of the problem.

\begin{figure}[t]
    \centering
    \includegraphics[width=\columnwidth]{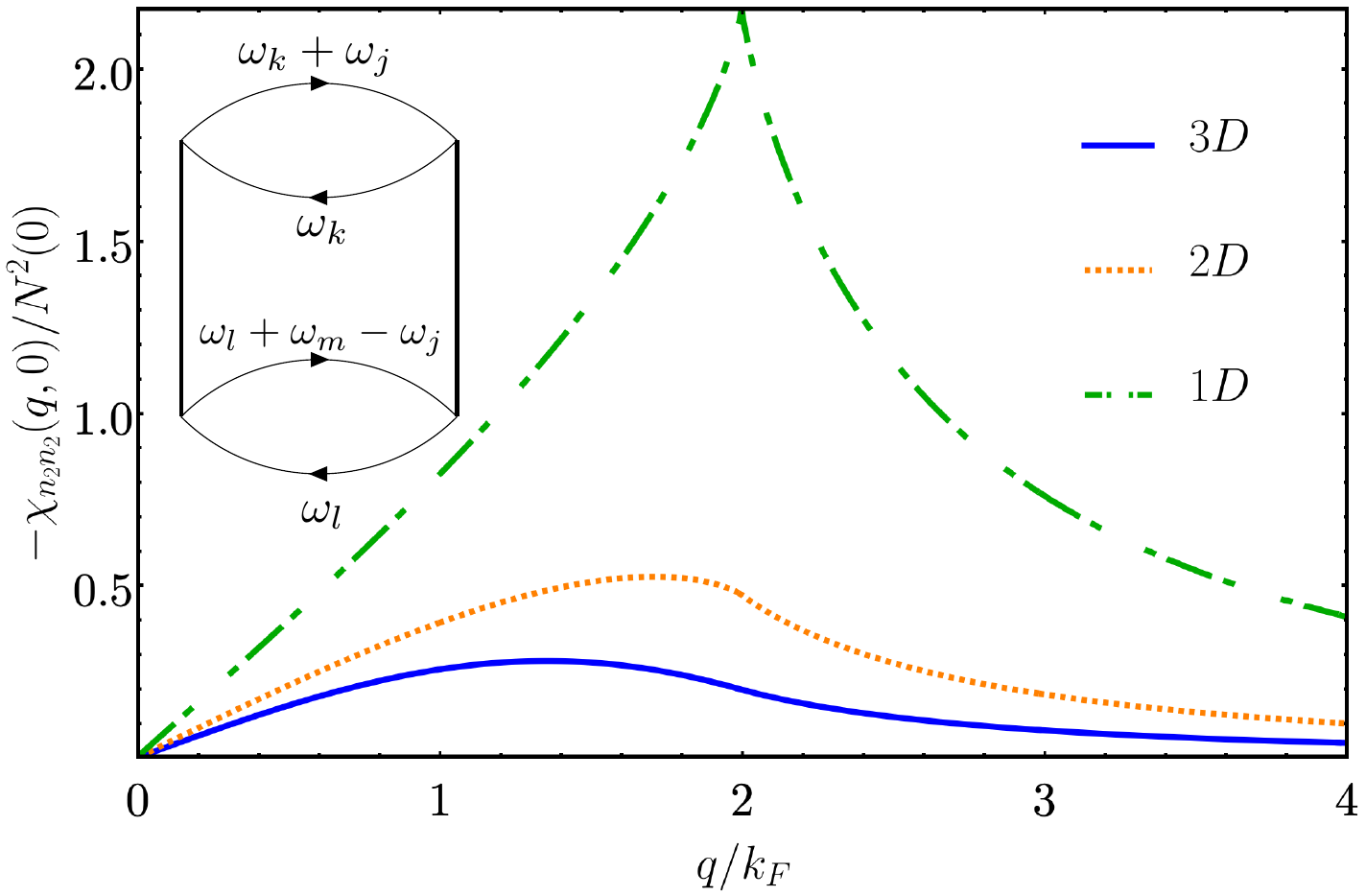}
    \caption{The fracton Lindhard function of a one-, two- and three-dimensional fracton gas in the static limit, in units of the squared single-particle DOS, $N(0)$, and as a function of momentum $q$ (in units of the Fermi momentum $k_{\rm F}$).
Inset: the Feynman diagram giving the dominant contribution to the fracton Lindhard function in the limit of large number of flavours. Thick lines represent extended vertices, $\omega_{l,k}$ are internal fermionic Matsubara frequencies, while $\omega_j$ is the bosonic frequency over which the convolution is performed.
}
    \label{Static}
\end{figure}

As shown above, the fracton density can be rewritten as the normal ordered product of two single particle densities. This fact greatly simplifies the evaluation of the non-interacting response functions for the fracton gas, since these can be recast (in the limit of large number of fermion flavors) in terms of the single particle responses of the electron gas~\cite{gv}.
Since a diagrammatic expansion cannot be defined directly for the causal response function, we will now consider the imaginary-time-ordered one~\cite{gv}, for which this task can be accomplished. 
We start by writing the temperature correlation function
\beq \label{eq:chi_T_time}
\chi^T_{n_2n_2}(\q,\tau) = - \braket{\mathcal T n_{\q}(\tau)n_{-\q}(\tau) n_{\q}(0) n_{-\q}(0)},
\eeq
where $\tau= i t$, $\mathcal T$ is the imaginary-time ordering operator~\cite{gv} and we expressed the two-particle density in term of single-particle one. In the limit of large number of fermion flavors, the response in Eq.~(\ref{eq:chi_T_time}) is dominated by diagrams with the maximum number of fermion bubbles. In frequency domain, these contributions give
\begin{equation} \label{eq36txt}
\chi^T_{n_2n_2}(\q,i\omega_m) = \frac{1}{\beta} \sum_m\chi^{T}_{nn}(\q,i\omega_j)\chi^{T}_{nn}(\q,i\omega_m-i\omega_j),
\end{equation}
where $\chi^{T}_{nn}(\q,i\omega_m)$ is the single-particle density-density response function, and $\omega_m$ and $\omega_j$ are bosonic Matsubara frequencies~\cite{gv}. 
Equation~(\ref{eq36}) can be represented diagrammatically as in figure~\ref{Static} (inset). There, the two Green's function bubbles represent the electron-gas Lindhard functions that appear on the right-hand side of (\ref{eq36}). The thick lines connecting the vertices of the bubbles define ``extended vertices'' (EV), and imply a convolution over the bosonic frequency $\omega_j$.
One can also define Feynman rules to construct more complex diagrams for the fracton density-density response function. Since Wick's theorem still holds as for a standard quantum gas, few modifications of the rules are needed to take into account that a density vertex is replaced by an extended one. In other words one can simply add the following two to the well known list of standard rules (see, e.g., Ref.~\cite{gv}).
\begin{itemize}
\item \emph{Identify lines connected by an EV}
\item \emph{Associate a frequency convolution to every EV}
\end{itemize}
%

Going back to Eq.~(\ref{eq36txt}), and using a well-known spectral representation for temperature correlation functions~\cite{gv}, the Sokhotski–Plemelj formula and Kramers–Kronig relations we can separate the real and imaginary parts of the susceptibility in the limit ($T \to 0$). We obtain
\barr \label{Lindhardtxt}
&& \chi'_{n_2n_2} (\omega) =  \int_{-\infty}^0 \frac{d \omega'}{\pi} \chi''_{nn}(\omega')
[\chi'_{nn}(\omega'_+) + \chi'_{nn}(\omega'_-)] 
\nonumber\\
&&
\chi''_{n_2n_2} (\omega) =- \int_{0}^\omega \frac{d \omega'}{\pi} \chi''_{nn}(\omega')
\chi''_{nn}(\omega-\omega'),
\earr
where $\chi'$ and $\chi''$ stand for the real and imaginary parts of a response function, respectively. Here, all functions depend on $\q$ and  $\omega'_\pm = \omega'\pm \omega$.
Together these expression represent the fracton Lindhard function and are the central result of our work. 


We first study the non-interacting (fracton) density-density response function, which we will name ``Lindhard function'' in analogy with the standard electron gas. Even in this simple case, the integrals in equation (\ref{Lindhardtxt}) can only be evaluated numerically by using the analytical expressions for the real and imaginary parts of the single particle Lindhard function $\chi_{nn}^{(0)}(\q,\omega)$ (which can be found, for example, in Refs.~\cite{gv,fetterval}). We start from the static limit [{\it i.e.}, $\chi_{n_2n_2}(\q,0)$ and evaluate Eq.~(\ref{Lindhardtxt}) in three, two and one dimensions.
The result is shown in figure \ref{Static}. At first sight the fracton Lindhard function looks qualitatively similar to its electron gas counterpart. However it is fundamentally different. Firstly, it is non-singular for $q=2 k_{\rm F}$, where $k_{\rm F}$ is the (fracton) Fermi momentum. 
In both two and three dimension the function is differentiable, while in one dimension is non-analytic but free from the logarithmic divergence of the 1D electron gas. 
Secondly, 
the fracton static response at $\q=0$ vanishes, in contrast to the electron-gas case where it approaches the density of states (DOS). This means that the fracton gas behaves  akin to an effectively gapped system. This can be understood by remembering that the long-wavelength static response gives a measure of the number of excited states available to the system for vanishing excitation energy. Since it is not possible to create excitations for vanishing  $\q$ without breaking dipole conservation, the response must vanish. In other words, the system does not couple to an homogeneous electric field.

\begin{figure}[t] 
    \begin{minipage}[b]{1\linewidth}
         \centering
         \begin{overpic}[width=\textwidth]{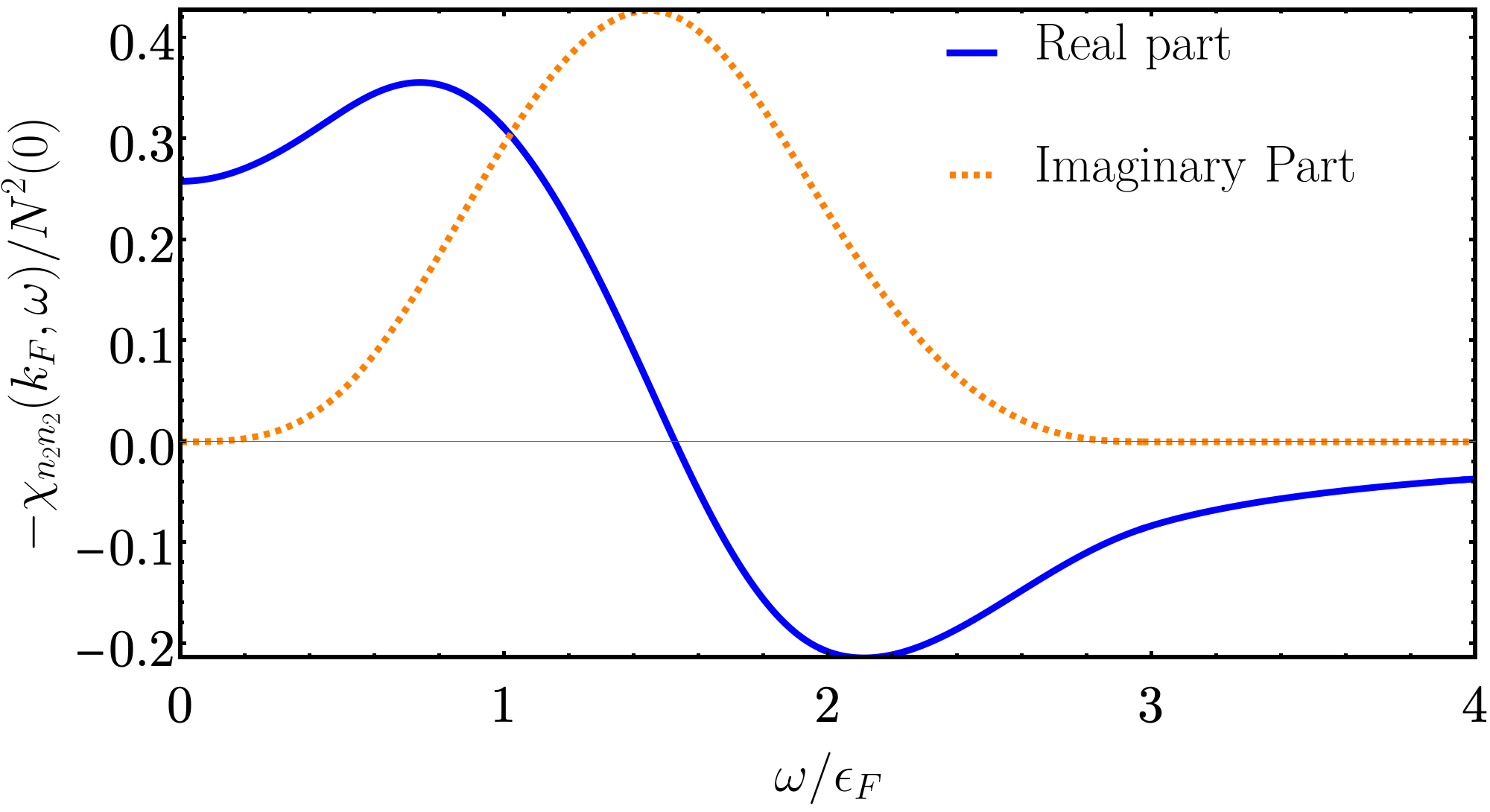}
         \put(0,1.3){(a)}
    \end{overpic}
     \end{minipage}\hfill
     \begin{minipage}[b]{1\linewidth}
         \centering
         \begin{overpic}[width=\textwidth]{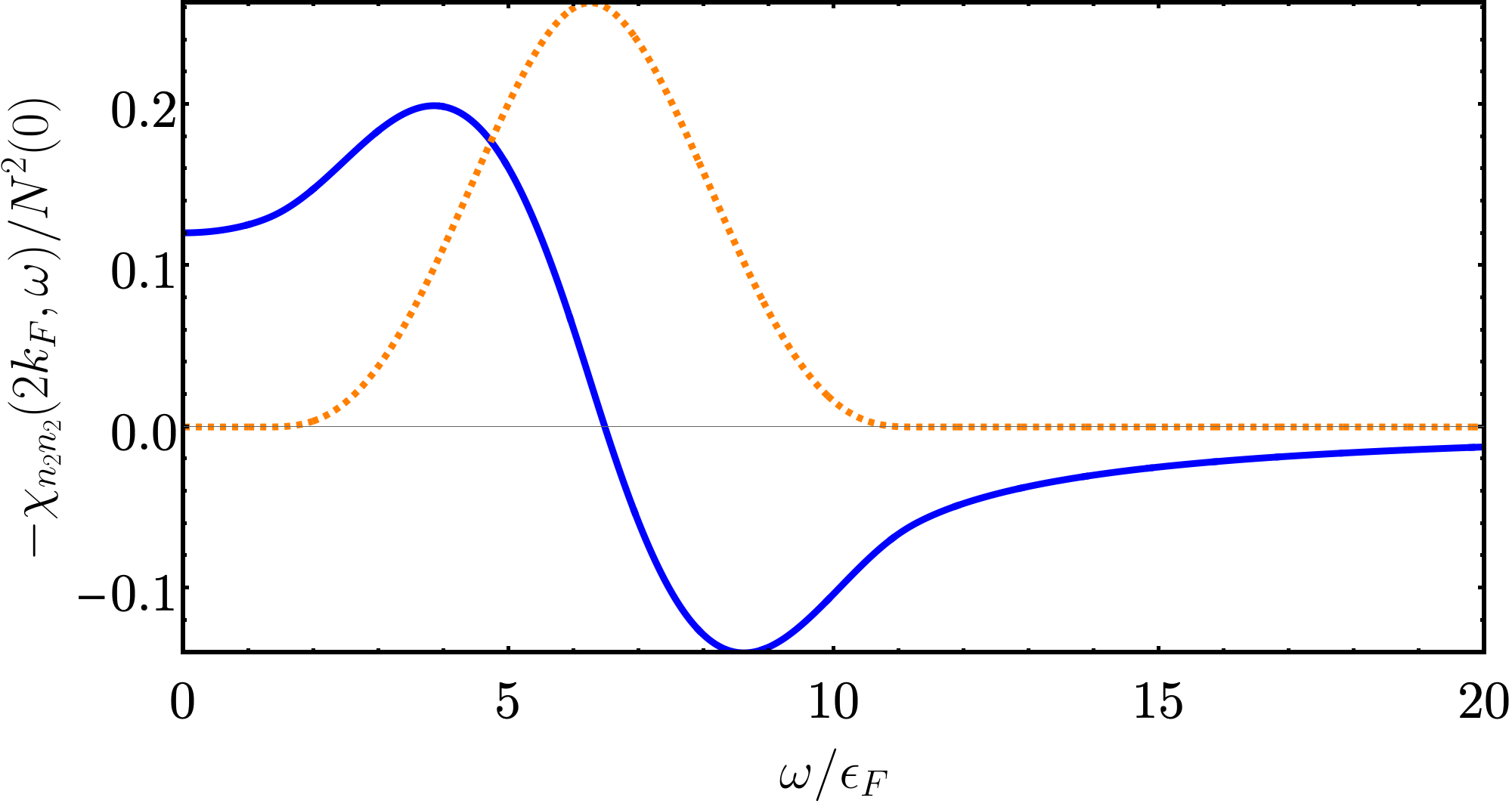}
         \put(0,0){(b)}
    \end{overpic}
     \end{minipage}\hfill
    \caption{Real and imaginary part of the density-density response function $-\chi_{nn}({\bm q},\omega)$ (in units of the squared single-particle DOS) of a three dimensional fracton gas as a function of the frequency $\omega$ in unit of single particle Fermi energy $\epsilon_{\rm F}$. (a) For $q = k_{\rm F}$. (b) For $q=2 k_{\rm F}$.}
    \label{dyn3d}
\end{figure}

We will now focus on the dependence of the Lindhard function on frequency, for fixed values of the momentum $\q$. 
In the case of the free electron gas, the imaginary part of the Lindhard function is non-zero only in a well defined range of frequencies, which is determined by the geometry of the Fermi surface and is given by the inequality $ \max{(0,\omega_{-}(q))} \leq \omega \leq \omega_+(q)$.
Here, $q=|\q|$ and $\omega_\pm(q)$ are obtained from the zeros of the denominator of the Lindhard function~\cite{gv}. These frequencies correspond to the minimum and maximum excitation energy available when an occupied state $\p$ is promoted to an empty state $\p+\q$.
In the fracton case, as shown e.g. in figure~\ref{dyn3d}, because of the smoothing effect of the convolution in Eq.~(\ref{Lindhardtxt}), the imaginary part of the Lindhard function does not vanish abruptly, though 
its spectral weight is still mostly concentrated between $\max{(0,\omega_{-}(q))}  \leq \omega \leq \omega_+(q)$.
We also find that the low-frequency behaviour of $\chi''_{n_2n_2} (\q,\omega)$ is different in the two regimes $q \le 2 k_{\rm F}$ and $q \ge 2 k_{\rm F}$.
For $q \le 2 k_{\rm F}$ [figure \ref{dyn3d} (a)], at small $\omega$ the tail of the imaginary part is a cubic function in the region $0\leq \omega \leq |\omega_{-}(q)|$, while it scales as $\omega^5$ for $\omega > |\omega_{-}(q)|$. Note that here, the point $\omega= |\omega_-|$ dividing the two regions is not a cusp as in the electron gas. This is again due to the smoothening effect of the convolution. On the contrary, for $q \ge 2 k_{\rm F}$ [figure \ref{dyn3d} (b)] only the $\omega^5$ scaling is present. These power laws can be obtained by computing analytically the convolution at fixed photon momentum by using the known expressions for the single electron response (see~\cite{Note1}). 

The behaviour of $\chi''_{n_2n_2} (\q,\omega)$ reflects onto that of the real part, which is connected to it via a Kramers-Kronig transform. The latter exhibits features, such as a swing in sign, that are much smoother than in the conventional electron gas \cite{gv}. We do not present the Lindhard function of two dimensional fractons, as 
differences and similarities with its electron gas counterpart are 
alike the 
ones we just presented for the
three dimensional case. 

%
Lastly we analyze the Lindhard function in the limit of small momenta, which can be extrapolated analytically from the integral in equation (\ref{Lindhardtxt}) to give (hereafter we reintroduce standard units)
\beq \label{fit}
\lim_{q \to 0}\frac{\chi_{n_2n_2}(q,\omega)}{N^2(0)L^D}= \hbar \frac{v^3_{\rm F}q^3}{2 \omega^2}.
\eeq
We notice that, akin to the electron gas, the limit is zero at $q=0$, but it scales with a higher power of the momentum, {\it i.e.} $q^3$ in lieu of $q^2$.
. 

Until now we have neglected the effect of interactions. We will now assume the constituent particles to be charged and to interact via a long-range Coulomb interactions of the form ${\hat H}_{\rm int} = \sum_{\bm q} v_q :{\hat n}_{\bm q} {\hat n}_{-{\bm q}}:$. Here, $v_q$ is the Fourier transform of the Coulomb interaction. The interaction Hamiltonian contributes~\cite{gv} to the imaginary-time evolution and to the definition of the thermal state in Eq.~(\ref{eq:chi_T_time}). Expanding the time-evolution operator and the thermal state order-by-order in $v_q$ and using Wick's theorem, one can define a diagrammatic expansion of the {\it interacting} fracton density-density response function. As usual, it is impossible to re-sum the entire series of diagrams: the simplest way to account for interactions is therefore to resum a subset of diagrams. We consider diagrams in which both density operators of a given interaction Hamiltonian are contracted with either of the two non-interacting bubbles of figure~\ref{Static} (inset), {\it i.e.} we avoid interactions running from one bubble to the other (see \cite{Note1} for the first few diagrams of this type).

 \begin{figure}[t]
    \centering
    \includegraphics[width=\columnwidth]{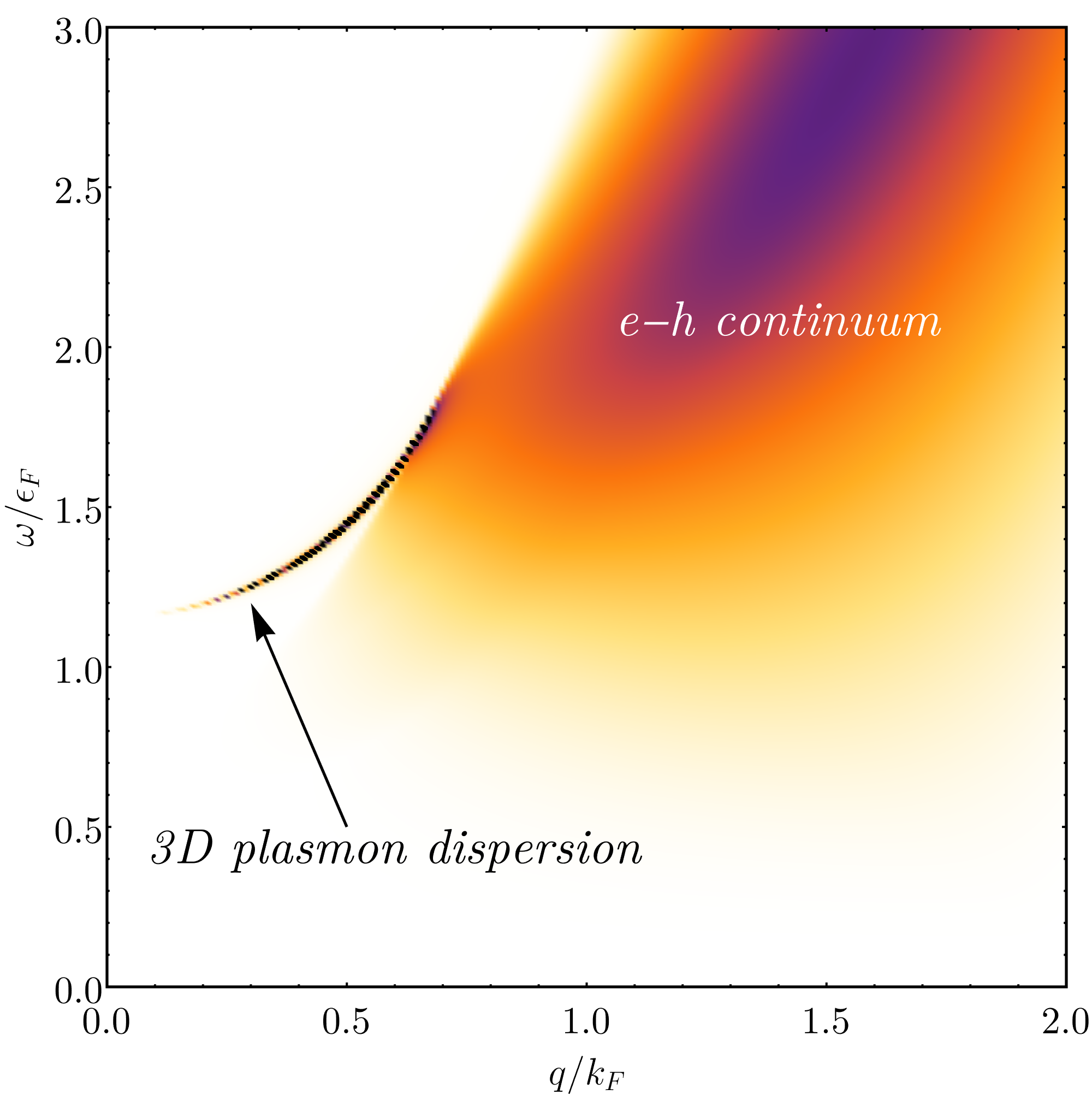}
    \caption{Density plot of the imaginary part of the RPA density response as a function of momentum $q$ (in units of $k_{\rm F}$) and frequency $\omega$ (in units of $\epsilon_{\rm F}$). The plasmon pole is clearly visible in the region outside the particle-hole continuum.}
    \label{RPA}
\end{figure}

By selecting only the diagrams with the largest number of bubbles, it is easy to show that this subset of diagrams can be written as a convolution of two single-particle random phase approximation (RPA) response functions. We therefore call this approximation the RPA expansion of the fracton density-density response function. Its expression is analogous to that of Eq.~(\ref{Lindhard}), where the response functions appearing on the right-hand side are given by $\chi_{nn}^{(0)}(\q,\omega)/\epsilon_{nn}^{(0)}(\q,\omega)$, where $\epsilon_{nn}^{(0)}(\q,\omega) = 1 - v_{q}\chi_{nn}^{(0)}(\q,\omega)$ is the single-particle dielectric function~\cite{gv}. 

By computing the imaginary part of the RPA response function we can study its poles and look at the collective mode of the system, which we call \emph{fracton-plasmons}. In figure \ref{RPA} we show the imaginary part of the RPA response in three dimensions where we highlight the plasmon dispersion. Qualitatively it is very similar to the case of the electron gas. However, it can be shown that in both three and two dimensions the \emph{fracton}-plasma frequency is double the electron-plasma frequency.
 
\section{Conclusions}
In this paper we have studied a simple model for the fracton quantum gas and its linear response to external fields. Paralleling the well known theory of the electron gas we have derived the density operator. Exploiting the duality with the translational invariant shell model we have argued that the quantisation scheme requires one to account for the natural constrains of the system. As we have shown, this reflects directly on the definition of the fundamental quantum operators. For example, the density operator is shown to act as a pair-density operator. 

Using this result, we have calculated the fracton density-density linear response functions.
Interestingly the restricted mobility of fracton excitations leads to a vanishing static response in the limit $q \to 0$. This is suggestive of an effectively gapped system and it is intimately related to the constraint. 
We have also shown that the high energy limit ($q \ll \omega$) of the response function scales with an higher power of the momentum compared to the response of an electron gas. This is expected to translates into different analytic properties of the impurity susceptibility \cite{gv} which, in turn, could be related to well-known sub-diffusivity of fracton systems \cite{glo,gln}. This problem, which requires further investigation, is beyond the scope of this work, which aims at laying down the general formalism. The formalism developed in this paper can help formulating the fracton physics with the powerful tools of many-body quantum field theory.

\emph{Acknowledgements}. We acknowledge support from the European Commission under the EU Horizon 2020 MSCA-RISE-2019 programme (project 873028 HYDROTRONICS) and of the Leverhulme Trust under the grant RPG-2019-363. This research was supported in part by grant no. NSF PHY-2309135 to the Kavli Institute for Theoretical Physics (KITP). We also thank Prof. Niels Walet for useful early discussions.

\appendix
\section{The  Kretzschmar transformation} \label{AppA}
In this appendix we review a canonical transformation of coordinates introduced by Kretzschmar \cite{kre} which recasts the translational invariant shell model (TISM) into a set of $N-1$ independent oscillators. 
We consider  the TISM Hamiltonian in matrix form
\beq
H = \P \Bigl( \mathbb I - \frac{1}N \mathbb E \Bigl ) \P + \R \Bigl( \mathbb I - \frac{1}N \mathbb E \Bigl ) \R
\eeq
where $\R=(\r_1,\dots,\r_N)$, $\P = (\p_1,\dots,\p_N)$, $\mathbb I$ is the $N \times N$ identity matrix, while $\mathbb E$ is a matrix whose elements are all equal to one. We also introduce the following linear canonical transformation for $\{\r_i,\p_i\}$
\beq \label{kt}
\begin{aligned}
\r_i &= \mathbb B \,  \r'_i,  \quad  \r'_N = \frac{1}{\sqrt N} \sum_i^N \r_i,\\
\p_i &= \mathbb B \,  \p'_i, \quad  \p'_N = \frac{1}{\sqrt N} \sum_i^N \p_i,
\end{aligned}
\eeq
where $\mathbb{B}$ is an orthogonal matrix which satisfies the following equation
\beq
\mathbb B^T \Bigl( \mathbb I - \frac{1}N \mathbb E \bigl ) \mathbb B = \left(
\begin{array}{cccc}
 1 & 0 & \dots & 0\\
 0 & \ddots & \dots & 0 \\
 \vdots & \dots &  1 &\vdots \\
 0 & \dots &  \dots & 0\\
\end{array} \right)
\eeq
In these new coordinates the Hamiltonian of the TISM reads 
\beq
H = \frac{1}2 \sum_{i=1}^{N-1} (\p_i'^2 + \r_i'^2).
\eeq
It may be tempting to use this new set of coordinates to compute the response of the system. This would only describe a free system of $N-1$ particles. The constraint the given by equation \ref{kt} would have to be enforced in the quantum states dynamics. 
\section{Eigenstates and operators of the dipole-conserving model} \label{AppB}
In terms of physical momenta of $N-1$ particles, the Hamiltonian is
\begin{equation}
{\hat H} = \sum_{j=1}^{N-1} \frac{{\hat {\bm \pi}}^2_j}{2} + \frac{1}{2} \left(\sum_{j=1}^{N-1} {\hat {\bm \pi}}_j\right)^2,
\end{equation}
where the last term is just ${\hat {\bm \pi}}_N$, which is fixed by the conservation of the physical center of mass momentum. It is easy to see that the coordinates conjugate to ${\bm \pi}_1, \ldots, {\bm \pi}_{N-1}$ are
\begin{equation} \label{eq:app_xi_def}
{\bm \xi}_j = {\bm r}_j + \sum_{k=1}^{N-1} {\bm r}_k, \quad j=1, \ldots, N-1,
\end{equation}
where ${\bm r}_j$ is the coordinate conjugate to ${\bm p}_j$. It is easy to see that ($i,j = 1, \ldots, N-1$)
\begin{eqnarray}
[{\bm \xi}_i, {\bm \pi}_j] &=& \left[ {\bm r}_i + \sum_{k=1}^{N-1} {\bm r}_k, {\bm p}_j - \frac{1}{N} \sum_{\ell=1}^{N} {\bm p}_\ell \right]
\nonumber\\
&=&
i\delta_{ij} - \frac{i}{N} + i - \frac{i}{N}(N-1)
\nonumber\\
&=&
i\delta_{ij},
\end{eqnarray}
where we used that $[{\bm r}_i,{\bm p}_j] = i\delta_{ij}$.

Since the Hamiltonian commutes with all ${\hat {\bm \pi}}_1, \ldots, {\hat {\bm \pi}}_{N-1}$, the many-body states can be labeled by the eigenvalues ${\bm \pi}_1, \ldots, {\bm \pi}_{N-1}$ of such operators. Many-body eigenvectors are denoted as $|{\bm \pi}_1, \ldots, {\bm \pi}_{N-1}\rangle$. This is just a shortcut for the usual occupation-number (or second-quantization) notation. In fact, instead of specifying the number of particle is each single-particle eigenstate (most of which are simply zero) we give the momenta for which the occupation number is equal to one. {\it I.e.}, one has occupations $n_{{\bm \pi}_1} = \ldots = n_{{\bm \pi}_{N-1}} = 1$ and zero otherwise. The state $|{\bm \pi}_1, \ldots, {\bm \pi}_{N-1}\rangle$ is just a Slater determinant of single particle wavefunctions $\phi_{\bm \pi}(\xi) = e^{i{\bm \pi}\cdot {\bm \xi}}$, {\it i.e.}
\begin{eqnarray} \label{eq:app_many_body_vectors_def}
&& |{\bm \pi}_1, \ldots, {\bm \pi}_{N-1}\rangle 
\nonumber\\
&&
= 
\frac{1}{\sqrt{N-1}} \sum_P (-1)^{S_P} \phi_{P{\bm \pi}_1}({\bm \xi}_1) \cdots \phi_{P{\bm \pi}_{N-1}}({\bm \xi}_{N-1}),
\nonumber\\
\end{eqnarray}
where $P$ is a permutation and $S_P$ its signature ($+1$ if it is a cyclic permutation of ${\bm \pi}_1, \ldots, {\bm \pi}_{N-1}$, $-1$ otherwise).

Taking the sum over $j$ of Eq.~(\ref{eq:app_xi_def}) we obtain the relation
\begin{equation}
\sum_{j=1}^{N-1} {\bm \xi}_j = N \sum_{k=1}^{N-1} {\bm r}_k,
\end{equation}
which we can use to invert Eq.~(\ref{eq:app_xi_def}) as
\begin{equation} \label{eq:app_r_xi_inverse}
{\bm r}_j = {\bm \xi}_j - \frac{1}{N} \sum_{k=1}^{N-1} {\bm \xi}_k.
\end{equation}
We can use Eq.~(\ref{eq:app_r_xi_inverse}) to derive the action of the density operator on the eigenstates of the system, proving that it is as described in the main text. Using Eq.~(\ref{eq:app_r_xi_inverse}) and the fact that $\phi_{\bm \pi}(\xi)$ is a plane wave,
\begin{eqnarray} \label{eq:app_density_action_slater}
&&
\sum_{j} e^{i{\bm q}\cdot{\bm r}_j} |{\bm \pi}_1, \ldots, {\bm \pi}_{N-1}\rangle 
= 
\frac{1}{\sqrt{N-1}} \sum_P (-1)^{S_P}
\nonumber\\
&&
\times
\sum_{j=1}^{N-1} \phi_{P{\bm \pi}_1-{\bm q}/N}({\bm \xi}_1) \cdots \phi_{P{\bm \pi}_{j} + {\bm q}(N-1)/N}({\bm \xi}_{j})
\nonumber\\
&&
\times 
\phi_{P{\bm \pi}_{j+1}-{\bm q}/N}({\bm \xi}_{j+1})
\cdots \phi_{P{\bm \pi}_{N-1}-{\bm q}/N}({\bm \xi}_{N-1})
\nonumber\\
&&
+
\frac{1}{\sqrt{N-1}} \sum_P (-1)^{S_P}
\nonumber\\
&&
\times
\phi_{P{\bm \pi}_1-{\bm q}/N}({\bm \xi}_1) \cdots  \phi_{P{\bm \pi}_{N-1}-{\bm q}/N}({\bm \xi}_{N-1})
.
\end{eqnarray}
The term on the last two lines stems from the action of $e^{i{\bm q}\cdot{\bm r}_N}$, and we used that, from Eq.~(\ref{eq:app_r_xi_inverse}),
\begin{equation} 
{\bm r}_N = -\sum_{j=1}^{N-1} {\bm \r}_j = -\frac{1}{N} \sum_{j=1}^{N-1} {\bm \xi}_j.
\end{equation}
It is immediate to see that, according to the definition~(\ref{eq:app_many_body_vectors_def}), such term reduces to the vector
\begin{eqnarray} 
|{\bm \pi}_1-{\bm q}/N, \ldots, {\bm \pi}_{N-1} - {\bm q}/N\rangle.
\end{eqnarray}
On the other hand, one can re-organize the the first three lines of Eq.~(\ref{eq:app_density_action_slater}) as follows. Collecting all the terms in which ${\bm \pi}_1$ has been shifted by ${\bm q}(N-1)/N$, one sees that their sum corresponds to the state
\begin{eqnarray} 
|{\bm \pi}_1+{\bm q}(N-1)/N, \ldots, {\bm \pi}_{N-1} - {\bm q}/N\rangle.
\end{eqnarray}
In a similar way one can reason for ${\bm \pi}_2,\ldots, {\bm \pi}_{N-1}$. Therefore, the first three lines correspond to the superposition
\begin{eqnarray} 
\sum_{j=1}^{N-1} 
|{\bm \pi}_1 - \frac{{\bm q}}{N}, \ldots, {\bm \pi}_j + \frac{{\bm q}(N-1)}{N} ,\ldots, {\bm \pi}_{N-1} - \frac{{\bm q}}{N}\rangle.
\nonumber\\
\end{eqnarray}

\section{The Lindhard functions of the dipole-conserving model}
The Lindhard function is written as~\cite{gv}
\barr \label{eq:app_Lindhard_def}
\chi_{nn}({\bm q},\omega) &=& \sum_{\{{\bm \pi}\}} \sum_{\{{\bm \pi}'\}} \frac{P_{\{{\bm \pi}\}} - P_{\{{\bm \pi}'\}}}{\omega +i\eta + E_{\{{\bm \pi}\}} - E_{\{{\bm \pi}'\}}}
\nonumber\\
&\times&
\big| \langle {\{{\bm \pi}'\}} | \sum_i e^{i{\bm q}\cdot{\bm r}_i} | {\{{\bm \pi}\}} \rangle \big|^2
,
\earr
where $\{{\bm \pi}\}$ is a short-hand notation for the combination of occupied momenta ${\bm \pi}_1, \ldots, {\bm \pi}_{N-1}$, $E_{\{{\bm \pi}\}} = (2m)^{-1} \sum_j {\bm \pi}_j^2$ is the energy of the many-body state, and $P_{\{{\bm \pi}\}} = e^{-\beta E_{\{{\bm \pi}\}}}/Z$. Here, $\beta$ is the inverse temperature and $Z$ the partition function. 

Note that in Eq.~(\ref{eq:app_Lindhard_def}) the product runs over all possible values of occupied momenta ${\bm \pi}_1, \ldots, {\bm \pi}_{N-1}$. It is then easy to see that summing over all values of occupied momenta is equivalent to summing over all many-body states with $N-1$ particles, {\it i.e.} over the (infinite) set of occupation numbers $\{n_{{\bm \pi}_\alpha}\}$ ($\alpha = 1, \ldots, \infty$) used to denote a many-body state and such that $\sum_\alpha n_{{\bm \pi}_\alpha} = N-1$. In this case, $P_{\{{\bm \pi}\}} \to Z^{-1} \prod_{\bm \pi} e^{-\beta n_{\bm \pi}(\varepsilon_{\bm \pi} - \mu)}$, where $\varepsilon_{\bm \pi} = {\bm \pi}^2/(2m)$ is the single-particle energy and $\mu$ is the chemical potential which is introduced to fix the total number of particles. Thus, $Z = \prod_{\bm \pi} \big(1+e^{-\beta (\varepsilon_{\bm \pi} - \mu)}\big)$

As explained in the main text, the action of $e^{i{\bm q}\cdot {\bm r}_i}$ on the eigenstates is to shift the momentum of particle $i$ by ${\bm q}(N-1)/N$, and the momentum of all other particles by $-{\bm q}/N$. Note that $i$ here is the index labelling a particle, not a value of the momentum. It then is possible to show that the operator $\sum_i e^{i{\bm q}\cdot{\bm r}_i}$ acting on the many-body state $| {\{{\bm \pi}\}} \rangle = |{\bm \pi}_1, \ldots, {\bm \pi}_{N-1}\rangle$ creates a superposition of states, each of which has one of the ${\bm \pi}$'s shifted by ${\bm q}(N-1)/N$ and all others by $-{\bm q}/N$ [see also Eq.~(\ref{eq:app_density_action_slater}) and the discussion thereafter], {\it i.e.}
\barr \label{eq:app_density_eigenstates}
&& 
\sum_{i=1}^N e^{i{\bm q}\cdot{\bm r}_i} | {\{{\bm \pi}\}} \rangle = |{\bm \pi}_1, \ldots, {\bm \pi}_{N-1}\rangle
\nonumber\\
&& 
= \sum_{j=1}^{N-1} 
|{\bm \pi}_1 - \frac{{\bm q}}{N}, \ldots, {\bm \pi}_j + \frac{{\bm q}(N-1)}{N} ,\ldots, {\bm \pi}_{N-1} - \frac{{\bm q}}{N}\rangle
\nonumber\\
&&
+ |{\bm \pi}_1 - \frac{{\bm q}}{N}, \ldots, , {\bm \pi}_{N-1} - \frac{{\bm q}}{N}\rangle
.
\earr
We remind the reader that the state on the last line of this equation is obtained by applying $e^{i{\bm q}\cdot{\bm r}_N}$ on the many-body state. This state contains the normalization factor $(N-1)^{-1/2}$ and, when it is introduced into Eq.~(\ref{eq:app_Lindhard_def}), it produces a term which is proportional to $(N-1)^{-1}$. This term, which cannot be rewritten in a simple way in terms of the electron-gas Lindhard function, vanishes in the thermodynamic limit and will be neglected in what follows.

Since the ${\bm \pi}$'s are dummy variables, the $N-1$ terms produced by plugging the second line of Eq.~(\ref{eq:app_density_eigenstates}) into the Lindhard function of Eq.~(\ref{eq:app_Lindhard_def}) can be rewritten as a single term in which ${\bm \pi}_1\to {\bm \pi}_1 +{\bm q} (N-1)/N$ and ${\bm \pi}_j \to {\bm \pi}_j - {\bm q}/N$ for $j\geq 2$. Their multiplicity produces a factor $N-1$ which exactly cancels the normalization factor $(N-1)^{-1}$, and therefore their contribution survives in the thermodynamic limit.

Eq.~(\ref{eq:app_density_eigenstates}) also implies that, for every choice of ${\bm \pi}_1, \ldots, {\bm \pi}_{N-1}$, the values of ${\bm \pi}_1', \ldots, {\bm \pi}_{N-1}'$ are completely determined, {\it i.e.} ${\bm \pi}_1' = {\bm \pi}_1 +{\bm q} (N-1)/N$ and ${\bm \pi}_j' = {\bm \pi}_j - {\bm q}/N$ for $j\geq 2$. The difference in energy in the denominator of Eq.~(\ref{eq:app_Lindhard_def}),
\barr
E_{\{{\bm \pi}\}} - E_{\{{\bm \pi}\}} &=& \sum_{j=1}^{N} \frac{{\bm \pi}^2_j}{2 m} - \sum_{j=2}^{N} \frac{({\bm \pi}_j-{\bm q}/N)^2}{2 m} 
\nonumber\\
&-&
\frac{1}{2m} \left({\bm \pi}_1 + {\bm q}\frac{N-1}{N}\right)^2
\nonumber\\
&=&
- \frac{{\bm \pi}_1\cdot{\bm q}}{m} - \frac{q^2}{2m} \frac{(N-1)^2+N}{N^2},
\earr
only depends on ${\bm \pi}_1$. 

We can then perform the sums over ${\bm \pi}_2, \ldots, {\bm \pi}_{N-1}$ keeping in mind that these must be kept different from ${\bm \pi}_1$ and ${\bm \pi}_1'$. As mentioned after Eq.~(\ref{eq:app_Lindhard_def}), this is equivalent to summing over occupation numbers $n_{{\bm \pi}}$ of $N-2$-particle states. This in turn implies that a vast majority of the factors in $Z$ are canceled, except $1+e^{-\beta (\varepsilon_{{\bm \pi}_1}-\mu)}$ in $P_{\{{\bm \pi}\}}$ and $1+e^{-\beta (\varepsilon_{{\bm \pi}_1'}-\mu)}$ in $P_{\{{\bm \pi}'\}}$. This results into
\barr \label{eq:app_Lindhard_def}
\chi_{nn}({\bm q},\omega) &=& \sum_{{\bm \pi}} \frac{f_{{\bm \pi}} - f_{{\bm \pi} + (N-1){\bm q}/N}}{\omega +i\eta + \varepsilon_{{\bm \pi}} - \varepsilon_{{\bm \pi} + (N-1){\bm q}/N}}
,
\earr
where $f_{\bm \pi}$ is the Fermi-Dirac distribution. Comparing with the usual formulas for the Lindhard function of an electron gas~\cite{gv}, $\chi_{nn}^{({\rm EG})}({\bm q},\omega)$, we see that
\barr \label{eq:app_chi_chiEG_relation}
\chi_{nn}({\bm q},\omega) = \chi_{nn}^{({\rm EG})}\Big({\bm q}\frac{N-1}{N},\omega\Big).
\earr
%
Note that this simple relation is due to the fact that we ignored the contribution to the Lindhard function coming from the last term of Eq.~(\ref{eq:app_density_eigenstates}), arguing that it vanishes in the thermodynamic limit. Thus, Eq.~(\ref{eq:app_chi_chiEG_relation}) is strictly valid only in the thermodynamic limit. It is then clear that, in this limit, the two functions coincide and all the fracton-like behaviour is lost. Therefore, a further constraint must be imposed if the latter should be retained, as it is done in the main text of this paper.

\section{The density-density response function}
In this section we provide a detailed calculation of the fracton density-density response which has been solely sketched in the main text. The momentum space time-ordered \emph{fracton}-correlation function is
\beq
\chi_{n_2n_2}(\q,t) = T\braket{n_{2,\q}(t) n_{2,-\q}(0)}.
\eeq
This expression can be factorised in a more user-friendly from as the product of single-particle correlation functions as 
\beq
\begin{split}
 T\braket{n_{2,\q}(t)n_{2,-\q}(0)}&=T\braket{:n_{\q}(t) n_{-\q}(t)n_{\q}(0) n_{-\q}(0):} \\&= \theta(t)\braket{:n_{\q}(t) n_{-\q}(t)n_{\q}(0) n_{-\q}(0):} \\& + \theta(-t) \braket{:n_{\q}(0) n_{-\q}(0)n_{\q}(t) n_{-\q}(t):}\\& =\chi_{nn}(\q,t)\chi_{nn}(\q,t).
\end{split}
\eeq
In frequency domain this gives 
\beq \label{C6}
\chi_{n_2n_2}(\q,\omega) = \int d\omega' \, \chi(\q,\omega') \chi(\q,\omega-\omega')
\eeq
As we have mentioned in the main text the diagrammatic evaluation of this expression is easier using the imaginary time/frequency finite temperature formalism. In the latter framework equation (\ref{C6}) can be written as 
\beq
\chi_{n_2n_2}(\q,\omega_n) = \frac{1}\beta\sum_{m=-\infty}^{+\infty} \chi(\q,\omega_m) \chi(\q,\omega_n-\omega_m)
\eeq
where $\omega_m$ is the bosonic Matsubara frequency.
Using the spectral representation for temperature correlation functions we can compute the causal response function in a more amicable form as \cite{gv}
\beq \label{eq37}
\begin{split}
\chi_{n_2n_2}(\q,\omega) = \frac{1}{\pi^2}&\int d\omega_1 d\omega_2\, \Im m \chi_{nn}(\q,\omega_1)\Im m\chi_{nn}(\q,\omega_2) 
\\&
\frac{n_B(\omega_2)-n_B(\omega_1)}{\omega + i \eta - \omega_1+\omega_2}.
\end{split}
\eeq
where $n_B(x)$ is the Bose-Einstein distribution and $\chi^{I}_{nn}(\q,\omega)$ is the single-particle density-density response function evaluated on the real-frequency axis~\cite{gv}. Resorting to the Sokhotski–Plemelj formula and Kramers–Kronig relations we can separate real and imaginary part and take the limit ($T \to 0$), we get
\beq \label{Lindhard}
\begin{split}
\begin{aligned}
\Re e \chi_{n_2n_2} (\q,\omega) &=  \int_{-\infty}^0 \frac{d \omega_1}{\pi} \, \Im m \chi_{nn}(\q,\omega_1)
\\ &\times 
[\Re e \chi_{nn}(\q,\omega+\omega_1) + \Re e \chi_{nn}(\omega-\omega_1)] \\
\Im m \chi_{n_2n_2} (\q,\omega) & =- \int_{0}^\omega \frac{d \omega_1}{\pi} \, \Im m \chi_{nn}(\q,\omega_1)
\\ &\times 
\Im m \chi_{nn}(\q,\omega-\omega_1). 
\end{aligned}
\end{split}
\eeq
For completeness, we can now show that the Lindhard function can be derived also at the zeroth order in the Feynman diagram expansion. According to the standard rules \cite{gv} at the lowest order of the finite temperature expansion, we have
\begin{widetext}
\beq \label{C8}
\chi_{n_2n_2}(\q,\omega_n) = \frac{1}\beta\sum_{m=-\infty}^{+\infty} \int d\p_1 d\p_2 \, \frac{1}{\beta^2}\sum_{jl}  \mathcal G^{(0)}(\epsilon_{\p_1},\omega_l)\mathcal G^{(0)}(\epsilon_{\p_1+\q},\omega_l+\omega_m)\mathcal G^{(0)}(\epsilon_{\p_2},\omega_j)\mathcal G^{(0)}(\epsilon_{\p_2-\q},\omega_j- \omega_m+\omega_n),
\eeq
\end{widetext}
where $\mathcal G^{(0)}(\epsilon_{\p_i},\omega_l)$ is the finite temperature free propagator. After performing the Matsubara summation in the internal fermionic frequencies equation (\ref{C8}) can be written as \cite{gv} 
\beq \label{eq36}
\begin{split}
\chi_{n_2n_2}(\q,\omega_n) =\frac{1}\beta\sum_{m=-\infty}^{+\infty}&\int d\p_1 d\p_2 \frac{n^F_{\p_1}-n^F_{\p_1+\q}}{\epsilon_{\p_1}-\epsilon_{\p_1+\q} + i\omega_m} 
\\ &
\frac{n^F_{\p_2}-n^F_{\p_2-\q}}{\epsilon_{\p_2}-\epsilon_{\p_2-\q} -i\omega_n + i\omega_m}
\end{split}
\eeq
where $n^F(\p)$ is the Fermi-Dirac distribution. Using again the spectral representation for temperature correlation functions we can rewrite equation (\ref{eq36}) in the form of equation (\ref{eq37}).  Finally we can again take the $T \to 0$ limit and separate real and imaginary parts. This, as expected, gives the fracton Lindhard function.
\section{Power law decay of $\Im m\,\chi_{n_2n_2}$} 
Here we show how to compute analytically the power low decay of the imaginary part of the Lindahrd function at fixed momentum. This can be done by considering the frequency scaling single particle imaginary response \cite{gv}
\beq
\begin{aligned}
    &\Im m \, \chi_{nn}(q\leq 2k_F,\omega) \simeq \omega \, \,\quad 0 \leq \omega \leq |\omega_{-}(\q)|, \\
    &\Im m \, \chi_{nn}(q\leq 2k_F,\omega) \simeq \omega^2 \quad  |\omega_{-}(\q)| \leq \omega \leq \omega_{+}(\q), \\
   & \Im m \, \chi_{nn}(q\geq 2k_F,\omega) \simeq \omega^2  \quad |\omega_{-}(\q)| \leq \omega \leq \omega_{+}(\q),
\end{aligned}
\eeq
where $\omega_{\pm}$ are the zeros of the denominator of $\chi_{nn}$. Using equation (\ref{Lindhard}) for the imaginary part we get
\beq 
\begin{aligned}
&\Im m \, \chi_{n_2n_2}(q\leq 2k_F,\omega) \simeq \omega^3 \,\quad 0 \leq \omega \leq |\omega_{-}(\q)|, \\
&\Im m \, \chi_{n_2n_2}(q\leq 2k_F,\omega) \simeq \omega^5 \,\quad  |\omega_{-}(\q)| \leq \omega \leq \omega_{+}(\q), \\
&\Im m \, \chi_{n_2n_2}(q\geq 2k_F,\omega) \simeq \omega^5 \,\quad  |\omega_{-}(\q)| \leq \omega \leq \omega_{+}(\q),
\end{aligned}
\eeq
where $\omega_{\pm}$ here are the zeros of the denominator of $\chi_{n_2n_2}$.
\onecolumngrid
\section{Random phase approximation diagrams}
In this section we show that akin to the case of the free electron gas it exists a subset of resummable diagrams which gives the RPA response. This is given by a convolution of single particle RPA response functions. This, in turn can be shown by considering the time evolution of the density-density response function
\beq
\chi_{n_2n_2}(\q,t) = -i\langle T [n_{\q}(t)n_{-\q}(t) n_{\q}n_{-\q}]S\rangle,
\eeq
where the scattering matrix $S$ is given by $S=T \exp{\bigl (-i \int_{-\infty}^\infty} H_I(t) dt \bigl)$ and we have omitted a denominator as we shall implicitly neglect disconnected diagrams \cite{gv}. The interaction Hamiltonian is $H_I= \sum_{\k} v_{\k} n_{\k} n_{-\k}$. We now consider the first order in the Dyson expansion of the scattering matrix, this gives 
\beq
\chi_{n_2n_2}(\q,t) = -\int dt_1 \langle T[n_{\q}(t) n_{-\q}(t)n_{\q} n_{-\q}\sum_{\k} v_{\k} n_{\k}(t_1)n_{-\k}(t_1)\rangle].
\eeq

 We now select Wick contractions of the above average such that each of the two density operators of the interaction Hamiltonian, $n_{\k}(t_1)$ and $n_{-{\k}}(t_1)$, is contracted with only another density operator. For example
the following density-density Wick contractions 
\beq
T\braket{n_{-\q}(t)n_{\q}}T\braket{n_{-\k}(t_1)n_{\q}(t)}T\braket{n_{-\q}n_{\k}(t_1)}
\eeq
gives
\beq
\chi_{n_2n_2}(\q,t) = \int dt_1 \chi_{nn}(\q,t) \chi_{nn}(\q,t_1-t) v_{\q} \chi_{nn}(\q,-t_1)
\eeq
where $\chi_{nn}(\q,t)= \sum_{\k} -i T\langle n_{\q}(t) n_{-\k}\rangle$ and conservation of momentum ensures $\k=\q$. Performing a frequency Fourier transform we get
\beq
\chi_{n_2n_2}(\q,\omega) = \int dt\Bigl( \int dt_1 \int d\omega' d\omega'' d\omega''' \chi_{nn}(\q,\omega') e^{i\omega't} \chi_{nn}(\q,\omega'')  e^{i\omega''(t_1-t)}\chi_{nn}(\q,\omega''') e^{-i\omega'''t_1}\Bigl)e^{-i\omega t},
\eeq
which gives the convolution
\beq \label{RPA1st}
\chi_{n_2n_2}(\q,\omega)= \int d\omega' \chi_{nn}(\q,\omega') v_{\q}\chi^2_{nn}(\q,\omega-\omega').
\eeq
Higher order calculations are only algebraically more involved but yield to the same result. The first few diagrams of this type are given in fig. \ref{RPA}. Fig. \ref{RPA}a shows the diagrammatic representation of equation (\ref{RPA1st}) while fig.  \ref{RPA}b the second order response.
\begin{figure}[H] 
    \centering
    \begin{minipage}[b]{1\textwidth}
         \centering
         \begin{overpic}[width=.5\columnwidth]{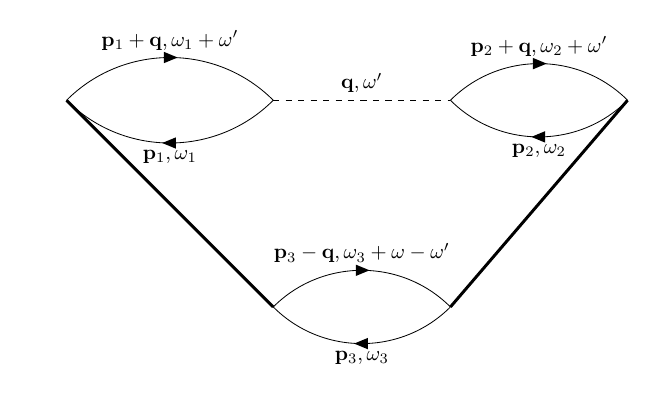}
         \put(0,1.3){(a)}
    \end{overpic}
     \end{minipage}\hfill
     \begin{minipage}[b]{1\textwidth}
         \centering
         \begin{overpic}[width=\columnwidth]{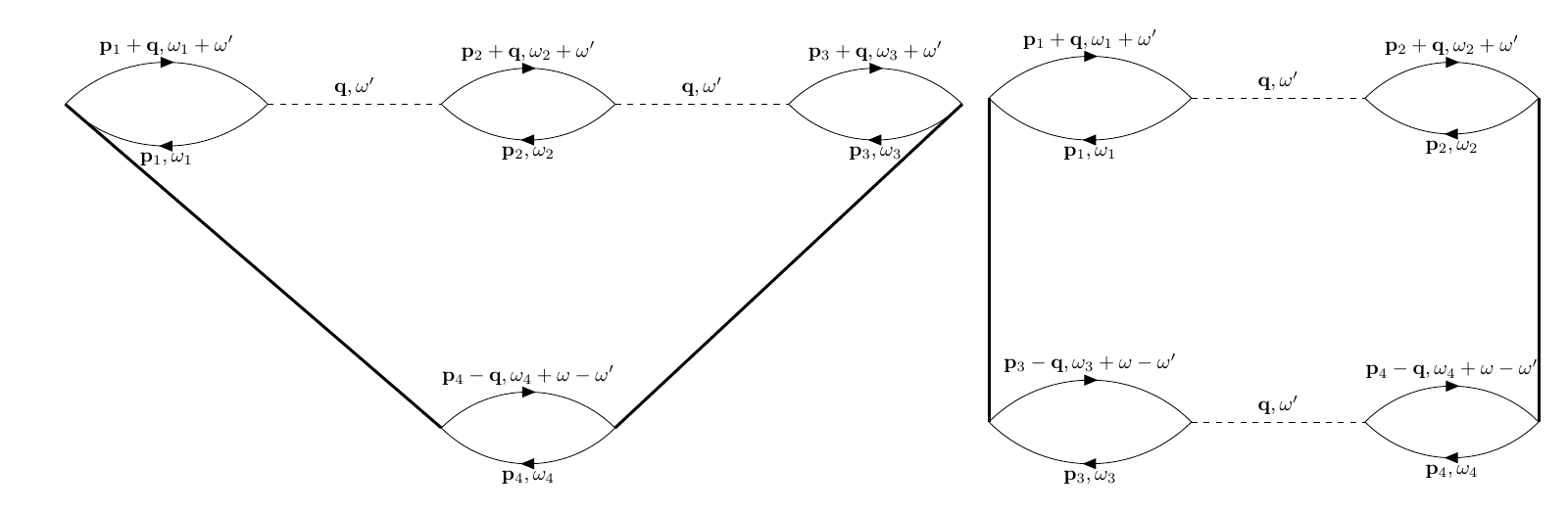}
         \put(0,0){(b)}
    \end{overpic}
     \end{minipage}\hfill
    \caption{ RPA response. (a) First order diagram with multiplicity 2. (b) Second order diagrams, the first with multiplicity 2.}
    \label{RPA}
\end{figure}
This series, considering the multiplicity of the diagrams given by the commutativity of the convolution (at each order there are $n+1$ diagrams, $n$ with multiplicity 2), can be resummed into
\beq
\chi^{RPA}_{n_2n_2}(\q,\omega) = \chi_{_{RPA}}(\q,\omega) * \chi_{_{RPA}}(\q,\omega).
\eeq
where ($*$) indicates the convolution product.
\twocolumngrid
\bibliography{mybib}
\end{document}